\documentclass{article}

\usepackage[utf8]{inputenc} 
\usepackage[T1]{fontenc}    
\usepackage{hyperref}       
\usepackage{url}            
\usepackage{booktabs}       
\usepackage{amsfonts}       
\usepackage{nicefrac}       
\usepackage{microtype}      
\usepackage{lipsum}		
\usepackage{graphicx}
\usepackage{url}
\usepackage{amsmath}
\usepackage{authblk}
\usepackage{hyperref}
\usepackage{doi}
\usepackage{fontenc}
\DeclareMathAlphabet{\mymathbb}{U}{BOONDOX-ds}{m}{n}
\usepackage{hyperref}

\usepackage{changepage}
\usepackage{comment}
\usepackage[normalem]{ulem}

\title{Doing data science with platforms crumbs: an investigation into fakes views on YouTube}

\author[1,*]{Maria Castaldo}

\author[1]{Paolo Frasca}

\author[2,4]{Tommaso Venturini}

\author[3]{Floriana Gargiulo}

\affil[1]{Univ.\ Grenoble Alpes, CNRS, Inria, Grenoble INP, GIPSA-lab, F-38000 Grenoble, France}

\affil[2]{CIS, CNRS, 59 Rue Pouchet, 75017 Paris, France}

\affil[3]{Gemass, CNRS,  59 Rue Pouchet, 75017 Paris, France}

\affil[4]{University of Geneva, Switzerland}

\affil[*]{Corrisponding Author: maria.castaldo@grenoble-inp.fr}
\date{}                     
\begin{document}
\maketitle
\begin{abstract}
This paper contributes to the ongoing discussions on the scholarly access to social media data, discussing a case where this access is barred despite its value for understanding and countering online disinformation and despite  the absence of privacy or copyright issues. Our study concerns YouTube's engagement metrics and, more specifically, the way in which the platform removes "fake views" (i.e., views considered as artificial or illegitimate by the platform). Working with one and a half year of data extracted from a thousand French YouTube channels, we show the massive extent of this phenomenon, which concerns the large majority of the channels and more than half the videos in our corpus. 
Our analysis indicates that most fakes news are corrected relatively late in the life of the videos and that the final view counts of the videos are not independent from the fake views they received. 
We discuss the potential harm that delays in corrections could produce in content diffusion: by inflating views counts, illegitimate views could make a video appear more popular than it is and unwarrantedly encourage its human and algorithmic recommendation.
Unfortunately, we cannot offer a definitive  assessment of this phenomenon, because YouTube provides no information on fake views in its API or interface. This paper is, therefore, also a call for greater transparency by YouTube and other online platforms about information that can have crucial implications for the quality of online public debate.
\end{abstract}


\maketitle
\section{Introduction}
\subsection*{Fake views, real trends}
    "We want to make sure that videos are viewed by actual humans and not computer programs"\cite{noauthor_how_nodate}. As stated in its official web pages, YouTube has a strong view count policy and has not been afraid of enforcing it. In December 2012, the platform deleted 2 billion views from the channels of record companies such as Universal and Sony \cite{gayle_youtube_2012} \cite{noauthor_youtube_2012} \cite{noauthor_fake_0500} \cite{noauthor_google_2014}. Over the years, countless youtubers have suffered sudden and drastic cuts to their views (and many have complained about it, often through YouTube videos). According to YouTube’s policies \cite{noauthor_fake_nodate} \cite{noauthor_keeping_nodate} \cite{noauthor_how_nodate}, these interventions aim at preserving a ``meaningful human interaction on the platform'' and to oppose ``anything that artificially increases the number of views, likes, comments or other metric either through the use of automatic systems or by serving up videos to unsuspecting viewers'' \cite{noauthor_fake_nodate}. 
    
Despite the media interest in the phenomenon \cite{kaminska_real-world_2015} \cite{noauthor_google_2015}, not much research has been carried out on the implementation of this policy. The general lack of studies on the subject is partly motivated by the fact that since 2016 YouTube has largely restricted access to its data through the API, making the researchers' task more difficult. Up to our knowledge, the only previous work concerning views count corrections is that of Marciel et al.\cite{marciel_understanding_2015} in 2016. This paper studies the phenomenon of views corrections in relation to video monetization, to identify possible frauds, drawing on research carried out on ads frauds in other social media \cite{chen_fake_2013} \cite{nagaraja_clicktok_2019}. In their work, Marciel et al.\ created some sample YouTube channels and inflated their views though bots. Strikingly, they found that ``YouTube monetizes (almost) all the fake views'' generated by the authors, while it ``detects them more accurately when videos are not monetized''. 

Although we consider this investigation into the correlation between monetization and views correction a first useful step toward understanding YouTube's policy, we believe that some other pressing questions should be addressed by the scientific community. For instance, can fake views have an impact on the success of a video and be used to manipulate YouTube's attention cycle? It is well known that, on social media, future visibility is highly dependent on past popularity, as trending contents tend to be favored by human influencers \cite{Rogers2018} and recommendation algorithms \cite{gillespie2016}, both of which are highly sensitive to trendiness metrics \cite{Venturini2019}. In YouTube in particular, the recommendation engine represents the most important source of views \cite{zhou_impact_2010} and, as admitted by its developers, ``in addition to the first-order effect of simply recommending new videos that users want to watch, [has] a critical secondary phenomenon of bootstrapping and propagating viral content'' \cite{covington_deep_2016}. Quite deliberately, YouTube's algorithm creates a positive feedback that skews visibility according to a rich-get-richer dynamic \cite{borghol_untold_2012} \cite{pinto_using_2013} \cite{szabogabor_predicting_2010}. As acknowledged by YouTube engineers: ``models trained using data generated from the current system will be biased, causing a feedback loop effect. How to effectively and efficiently learn to reduce such biases is an open question'' \cite{zhao_recommending_2019}. 

This is where fake views come into play. Indeed, if the correction of illegitimate views happens too late, these views have the potential to weight in the cycle of trendiness \cite{Castaldo2020} and unfairly propel their targets. If YouTube fake views correction is significantly slower than its recommendation dynamics, then artificially promoted videos risk to be favored by human and algorithmic recommendations, and thus reach larger audiences and collect extra real views. If, before being deleted, fake views are able to trigger a cascade effect that increases the visibility of some content, then they may be used to manipulate online debate. Not unlike social bots \cite{ferrara_rise_2016} and paid commentators \cite{King2017}, fake views could give the false impression that some content is highly popular and endorsed by many, thus distorting public debate and ultimately endangering democratic processes \cite{zhang_impact_2013} \cite{ratkiewicz_detecting_nodate} \cite{metaxas_obscurity_nodate} \cite{bessi_social_2016} \cite{llewellyn_for_2018} \cite{bastos_brexit_2017} \cite{canada} \cite{lipton_perfect_2016}.

\subsection*{Approach and tentative results}
While much research has been dedicated to identifying social bots \cite{ratkiewicz_truthy_2011} \cite{boshmaf_key_nodate} \cite{soldo_optimal_2012} \cite{venkataraman_tracking_nodate} and analyzing artificially produced contents \cite{spammingBotnets} \cite{ferrara_bots_2020} \cite{yang_how_2017} \cite{boshmaf_design_nodate} \cite{xie_spamming_nodate} \cite{chen_spatial-temporal_2008} \cite{correia_statistical_2012} \cite{subrahmanian_darpa_2016} \cite{lee_uncovering_2010}, no attention has been devoted to the role played by fake views in content diffusion. This work addresses this open and urgent question by analysing YouTube data at an unprecedented temporal granularity. As detailed below, we collected two datasets from more than a  thousand French `politics and news' channels. The first dataset is a 17 month collection of the hourly views counts of over 270.000 videos. The second dataset collects the views evolution of a thousand videos, but does so every five minutes. These datasets and their combined analysis allow us to examine, for the first time, the timing of YouTube fake views corrections and to rise concern about their consequences.

In summary, our analysis has lead to three main findings. Our first finding concerns the remarkable size of the phenomenon, since we identify fake views corrections in almost all the channels of our larger corpus and more than a half of the videos. Our second finding is that the rhythms of fake views corrections are inconsistent with those of view production, thereby suggesting that significant delays could occur between the generation of fake views and their correction.
Our third finding is the existence of a correlation between fake and real views: videos with more fake views also collect a larger number of legitimate views. 

\subsection{Disclaimer}
Despite our best efforts, our analysis cannot provide a definitive proof of the influence of fake views, because YouTube offers very little temporal information about fake and real views. At any given moment, we can retrieve the total number of views collected until then by a video, but not the history of their evolution -- forcing us to collect the views count at regular intervals.
Even more frustrating, through its interface or API, YouTube offer no information about fake views removal. This information, it is worth pointing out, does not raise any particular security or privacy issue -- if the real views count is revealed, why not the fake views one? The fake views count would provide a crucial cue for fact-checkers, journalists and scholars to identify possible shady operations around (although not necessarily by) a video or a channel, and yet YouTube decides to hide this information, which it could easily make available. As many of its likes, the platform share its data only to the extent that they help its business model. Like it or not, however, online platforms have become more than commercial players and the influence they have acquired on public debate should bind them to greater transparency.
We wish we could relegate our methodological troubles to a footnote and focus on the results. Unfortunately, YouTube does not make this possible. This paper thus is also a way to share with the research community a few of workarounds we developed to study attention cycles despite YouTube's opacity and in particular a method to reconstruct fake views count through frequent monitoring and machine learning techniques.

\section{Results}
\label{sec2}

Our results are mainly based on the analysis of a data set which records \textit{hour by hour} and for 17 months the number of the views collected by 270.133 videos published by 1.064 French channels dealing with news and politics. The limitations imposed by YouTube API made it impossible to collect data more frequently (or to monitor videos more than a week after their publication). The hourly frequency, of course, leads to a significant loss of information. As no direct information is provided about views corrections, we can
only \textit{estimate} them by observing the hours with a negative delta in the views count, i.e. the hours in which a video loses more views than it gains. 
In the hours in which the number of new views exceeds the number of removed views, the platform's interventions become invisible. For this reason, we did not draw our conclusions from the direct analysis of this dataset: instead, we have developed a machine learning method to discover these hidden correction. This method is described in details in the Data and Methods section.

In the Data and Methods section, we first give a quantitative estimation of the amount of information lost due to the hourly frequency of collection, in comparison with a 5-minute frequency.
Then, we propose two methods to reconstruct the number of YouTube corrections (a Benchmark Method, based on an heuristic procedure, and the Reconstruction model, a machine learning algorithm) and we compare their performance. The Reconstruction model, allowing to better reduce the information loss, is finally applied on all the original time series to reconstruct the number of views removed by YouTube but not directly visible in the raw data set.
In the rest of this section, we present the results obtained from the analysis of the data reconstructed in this way.

\begin{figure}
    \centering
    \includegraphics[width=\textwidth]{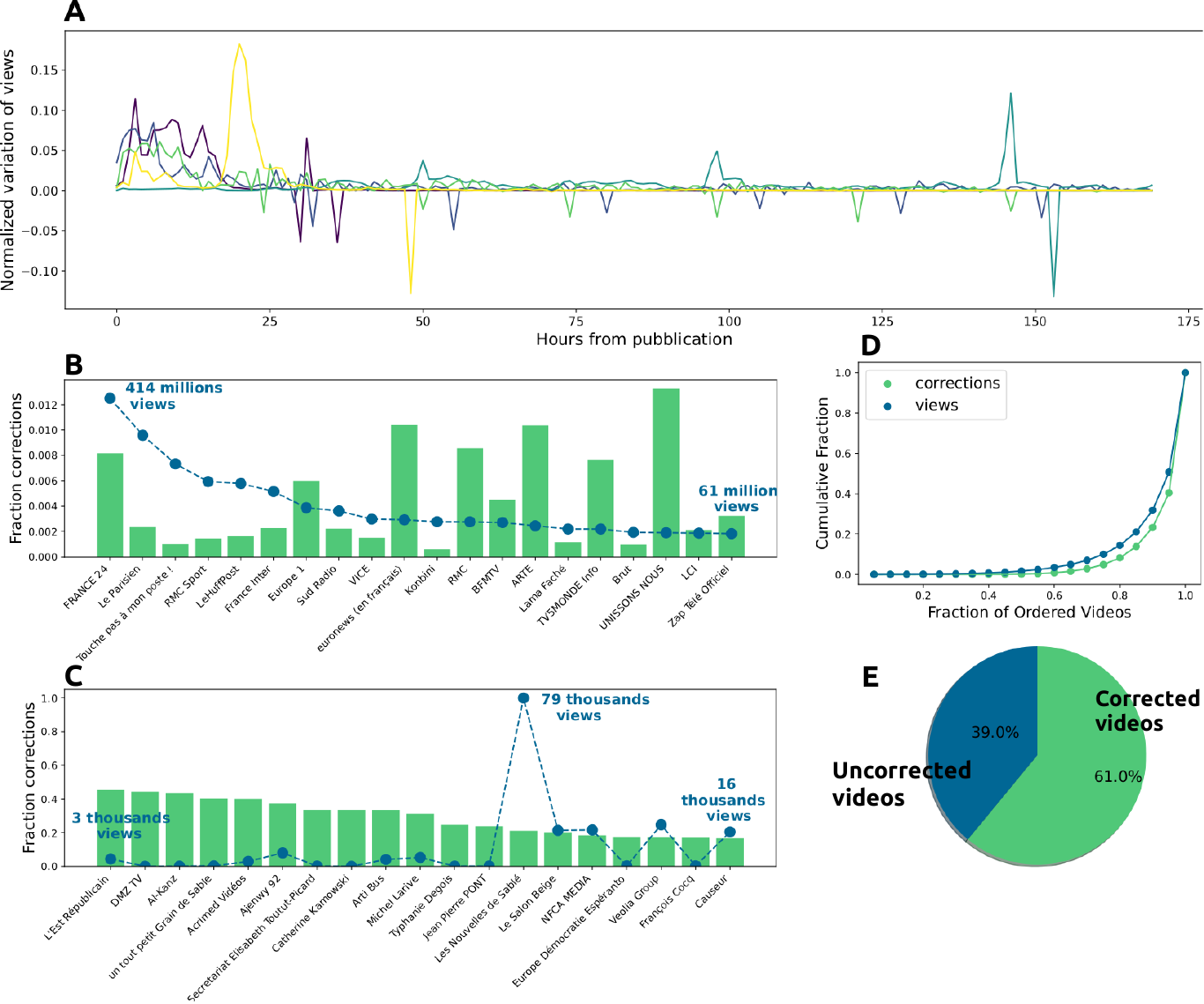}
    \caption{\textbf{A:} 5 sample videos and their hourly evolution of views. \textbf{B:} The 20 most viewed channels in our dataset and their fraction of corrections over real views. \textbf{C:} The most corrected channels in terms of fraction of corrections over real views. \textbf{D:} Lorenz curve of the distribution of views and corrections among different videos \textbf{E:} Percentage of videos concerned by the policy.}
    \label{fig:fig1}
\end{figure}
\subsection{Scale of the phenomenon}
The removal of false views is evident when the series of hourly views has negative entries: some examples are shown in Figure~\ref{fig:fig1}A. In fact, we have found out that fake views corrections are extremely common: we detected corrections for almost all monitored channels (90\% of them) and for 61\% of the videos in our corpus. Such a large scope underscores the importance of better understanding how these corrections are made. 
In fact, corrections in our corpus amount to about 22.5 millions. Although they represent, on average, a seemingly modest 0.5 percent of the total views, their number remains impressive and, more importantly, their distribution is very uneven. If we look at the Lorenz curve (Figure~\ref{fig:fig1}D) of the distribution of corrections among different videos, we can see that most of the corrections (more than 80\%) are concentrated on only 20\% of the videos. 
In comparison, the concentration of corrections appears to be stronger than that of legitimate views.

The heterogeneity of corrections is confirmed when we examine the most popular and the most corrected videos. 
Figure~\ref{fig:fig1}B shows the 20 most popular channels in our dataset and the percentage of actual views that corrections account for. These very popular channels, which are mainly traditional media channels such as TV stations, newspapers, and radio stations, still show marked differences in their corrections (between 0.1 percent and 1.3 percent).  In contrast, if we look at the 20 channels with the highest fraction of corrections to actual views (Figure~\ref{fig:fig1}C), we find channels with 40 percent corrections. These channels, which are mostly platform-native youtubers, collect substantially fewer views than the top channels (averaging about half a million views over the collection period).

\subsection{Correction rhythms}

\begin{figure}[h]
    \centering
    \includegraphics[width=\textwidth]{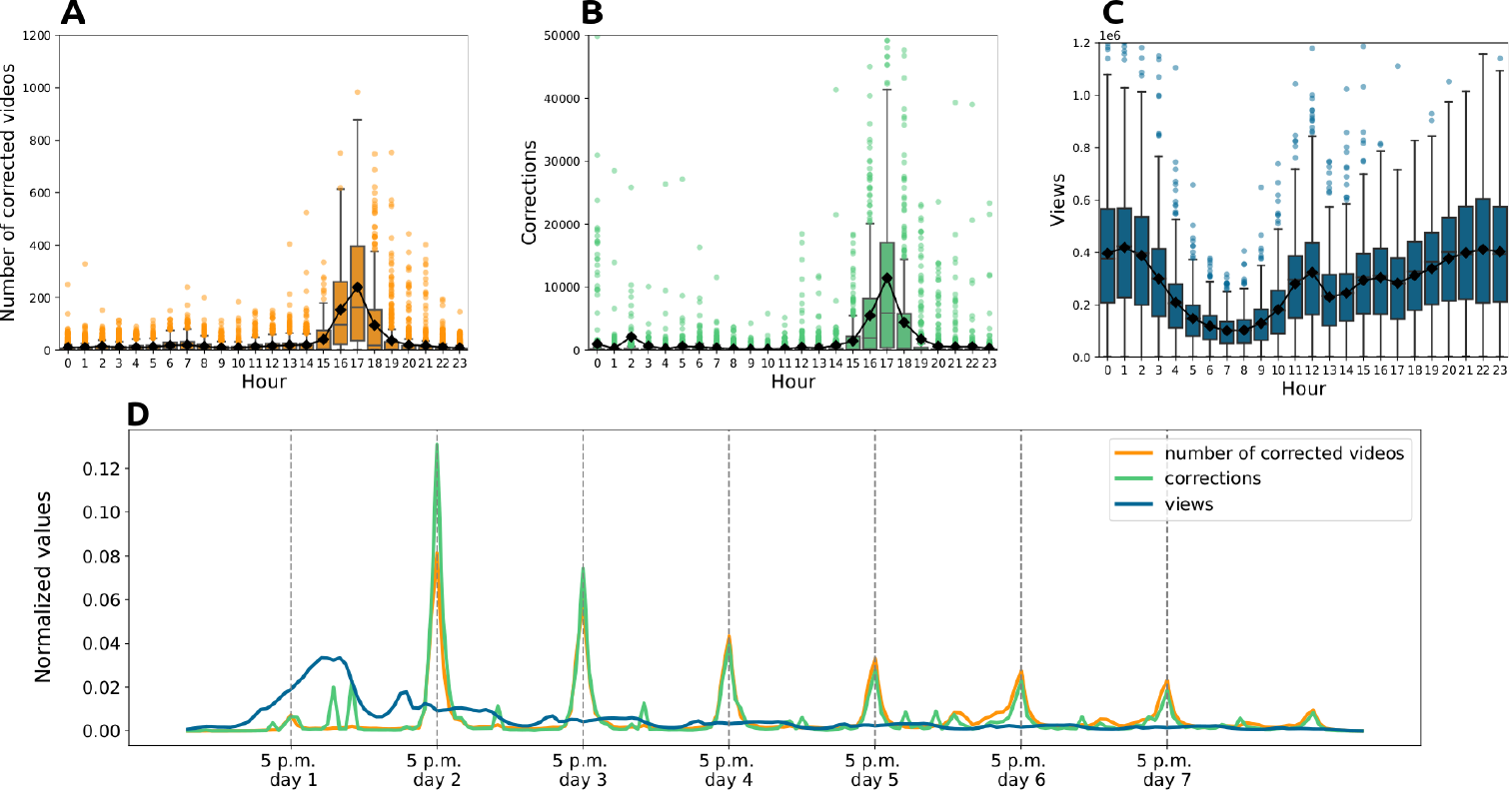}
    \caption{\textbf{A:} Distribution of the number of corrected videos per hour of the day. \textbf{B:} Distribution of corrections per hour of the day.
    \textbf{C:} Distribution of views per hour of the day. \textbf{D:} Normalized number of corrected videos, corrections and views since midnight before publication.}
    \label{fig:fig2}
\end{figure}

The view correction activities by YouTube have some interesting recurrences. If we look at Figure~\ref{fig:fig2}B, we can see how corrections are distributed according to the time of day. We can see, for example, that while the median number of removed views  hovers around a few dozen at most hours, it rises to more than 10,000 at 5 p.m. and hovers around 5,000 at 4 and 6 p.m.
This observation is confirmed if we study the number of videos corrected at each hour of the day (Figure~\ref{fig:fig2}C). Again, while normally the median number of corrected videos is close to zero, between 4 p.m. and 6 p.m. it is around 150 to 250. These rhythms of correction are peculiar and completely different from the rhythms at which views are made on the platform. In fact, if we look at Figure~\ref{fig:fig2}C, we can see how the views are distributed at different times of the day: they present a completely different behavior, with a minimum of hourly views around 7 a.m.\ and peaks of views in the evening or night hours (9 p.m.\ to 2 a.m.).

In summary, views are distributed quite evenly during the day, according to circadian rhythms \cite{castaldo_covid}, while corrections are concentrated in specific time slots. This fact suggests that most correction activities take place once a day, every 24 hours. This rhythm not only is unrelated to the rhythms of views production, but also seems rather slow given the fast pace at which content is propagated on the platform and the speed at which suggestions from recommendation systems are updated. For comparison, most of the videos in YouTube's "trending" section\footnote{https://www.youtube.com/feed/trending} are less than 24 hours old.\footnote{After collecting the top 20 videos in the trendy section in 7 different days, we found out that only the 25\% have been published since more than a day.} Therefore it is legitimate to ask whether such a frequency of corrections is too low to prevent possible interference of non-legit views with human and algorithmic recommendation.

If instead of looking at the distribution by time of day we study how corrections and views are distributed starting from publication, another interesting result emerges.
In Figure~\ref{fig:fig2}D, on the x-axis, we considered a time starting from midnight prior to the publication of a video. For each hour starting from midnight prior to publication, we summed the number of views, corrections, and corrected videos (with the convention of considering these quantities zero prior to publication). The expedient of measuring time from the midnight before publication allows us to maintain the periodicity of the corrections phenomenon, which would be lost if we summed the time series from their time of publication. 
What can be seen in this graph (normalized to make corrections and views comparable) is that views are much more concentrated in the first few hours after a video is published, while corrections are more spread out over the life of the videos. In fact, most views occur before 5 p.m. on the second day, the time when substantial corrections begin to occur.  This apparent delay in the corrections, together with their low frequency, prompt us to investigate more deeply the relationship between corrections and popularity.

\subsection{Late corrections and popularity}
To gain a deeper insight into correction mechanisms and investigate possible interference with the recommendation system, we should consider the timing of corrections not only in absolute terms, as done in the previous section, but also relative to when videos collect the most views. In particular, it is crucial to ask whether corrections are made before or after a video reaches its peak popularity. If corrections occur after this peak, one might suspect that by inflating the number of views, illegitimate views could make a video appear more popular and unfairly boost its human and algorithmic recommendation. 

Figure~\ref{fig:fig2}(left) presents the percentage of corrections made before videos reach a certain percentage of real views. For each video, we calculated the percentage of views collected at each hour. We then counted the number of fake views that occur before a certain percentage of real views. As the graph shows, most of the illegitimate views are corrected after the videos have garnered most of the real views. On average, only about 10 percent of the corrections are made before the videos have achieved 80 percent of the views. Even more striking is the fact that as many as 54 percent of corrections are made after the videos have stopped collecting real views, at the end of their lives. It is therefore quite clear that most illegitimate views are removed very late in the lives of most videos, well after their popularity has peaked and begun to decline.  

\begin{figure}[h]
    \centering
    \includegraphics[width=\textwidth]{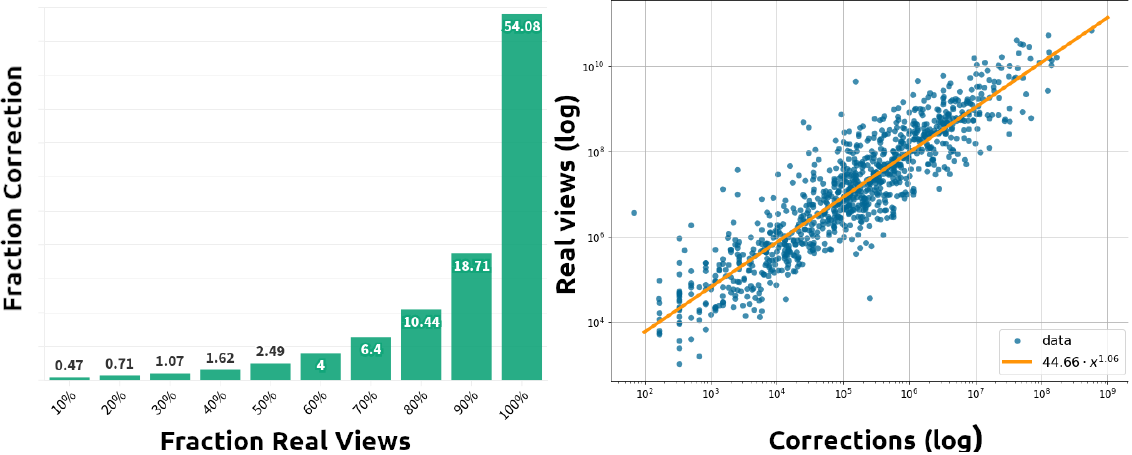}
    \caption{\textbf{Left:} Fraction of corrections occurring after different percentages of real views. \textbf{Right:} Correlation between real and fake views per channel.}
    \label{fig:fig3}
\end{figure}

In order to understand whether these delays have an effect on video popularity, we need to better investigate the relationship between fake views and popularity. In Figure~\ref{fig:fig3}(right) we examine the relationship between the number of fake views and the total number of legitimate views per channel.  In principle, these two quantities should be independent: in fact, if YouTube's correction policy were sufficiently fast and efficient, illegitimate views should have no impact on real views and, at the same time, there is no apparent reason why more popular videos should attract more fake views than others. Yet, Figure~\ref{fig:fig3}(right) shows a strong linear correlation between the logarithms of the two quantities. With a R-square equal to 0.819, the logarithms of fake and real views are related by a linear regression with intercept equal to 1.7704 and slope equal to 1.0574. The p-values associated with these quantities, as the reader might expect by looking at the plot, are very low, specifically less than 0.001. This linear correlation between the logarithms of the two quantities results in the following relationship between corrections ($c$) and real views ($v$):
\[
c = 58.94 \cdot v ^{1.06}
\]
The relationship between these two quantities is hence slightly more than linear, with a $95\%$ confidence interval on the exponent being within $[1.026, \; 1.089]$.

In view of this clear correlation, it becomes natural to wonder about a possible causal relationship between fake and real views. In particular, we would like to know whether fake views have an impact on popularity or not. Unfortunately, with the information we have, we are unable to answer this question. 
Although we have been able to reconstruct the time series of the \textit{removal} of fake views, we are unable to gather any information on the timing of \textit{generation} of these fake views. Without knowing when fake views are made, we cannot in any way know whether they are recorded before or after the videos become popular, and thus we cannot investigate the \textit{causal order} of the two phenomena. This information is not provided by the platform in any form and we hence cannot investigate the causal relationship between fake views and popularity.

\section{Data and Methods}\label{sec3}

\subsection{Data}\label{subsec2}
For our analysis, as described above, we used a large dataset with hourly frequency, which we will refer to as the \textit{hourly dataset}. We also collected a smaller dataset with 5 minute frequency, which we refer to as the \textit{5 minute dataset}: its purpose is to assess how much information was lost by collecting the time series with hourly frequency and to test our methods to reconstruct the lost information. The details of the two datasets are presented next.

\subsubsection*{Hourly dataset}\label{subsubsec2}
Starting from January 2021, through a collaboration with the Qatar Computing Research Institute (QCRI), we collected the time evolution of views of any video published by a list of over a thousand French YouTube channels dealing with news and politics. This dataset is particularly interesting as it cannot be straightforwardly obtained through the official YouTube application programming interface (API) \cite{noauthor_api_nodate}. Indeed for the latest several years the YouTube API \cite{noauthor_api_nodate} has restricted the accessible engagement metrics of a video to the current values only, without providing their temporal evolution any longer: this restriction complicates temporal studies on development and prediction of online attention and indeed studies on the temporal evolution of engagement metrics have not been carried out since 2014 \cite{szabogabor_predicting_2010} \cite{pinto_using_2013}. 

The  dataset  covers  1064  popular  French  channels  that  we deemed influential  in  the  French  public  debate. These  channels,  with  their  description,  are  available at \cite{data1}.  The channels have been selected through a qualitative analysis of the French YouTube, aiming to identify relevant actors that diffuse political opinions through the platform.  The selected channels belong to the following categories: local and national media; influential YouTubers discussing political topics; militant associations; politicians; political parties; Yellow Vests groups; associations devoted to public causes; large public or private institutions. YouTube provides no information about the location from which videos are viewed, but since the channels of our corpus focus on French public debate, we can assume most of their viewers to be based in France. 

For these channels, we recorded hour by hour the evolution of views of each video published after the 1st of January 2021, for an entire week after publication (170 hours).  Between the 1st of January 2021 and the 10 of May 2022,  we collected the views time series of 270.133 videos. The dataset is available at \cite{data2} (resource IDs have been mapped to anonymize values consistently along the dataset).  The choice of collecting only one week of views evolution is justified by noticing that news channels often collect the majority of their views in few days after publication, presenting a strong initial burst followed by a power-law decay \cite{yu}. Indeed our data confirm this fast decrease of users engagement as only $3\%$ of views is obtained in the last 24 hours.
The choice to collect views every hour was dictated by the constraints on YouTube's API, which does not allow more than 10,000 requests per day with the same access key. As a result, wanting to monitor the French news and political sphere on YouTube, and hence a certain amount of videos every day, we were forced to collect the data with this frequency. 

\subsubsection*{5-minutes dataset}\label{subsubsec2}

To assess the amount of information lost by collecting data every hour, we collected a smaller dataset with a 5 minutes frequency. This frequency has been chosen so to minimize the information loss and is in practice the highest useful frequency of data collection, since we have empirically observed that view counts are updated no faster than every 5 minutes. This dataset contains 1012 videos posted between February 2, 2022 and February 16, 2022. The dataset is available at \cite{data3}. Video identifiers have been anonymized consistently with the hourly dataset.

\subsection{Method to estimate corrections}\label{subsec2}
To reduce the loss of information due to collecting data on an hourly basis, we have devised a method to reconstruct the original corrections and interventions. To remove any bias possibly caused by the hourly aggregation, we need to infer the \textit{real} corrections $c_h^i$ made by the platform at hour $h$ on video $i$ from the observable quantity $\tilde v_h^i = v_h^i - c_h^i$, where $v_h^i$ are the effective non-observable views collected by video $i$ during hour $h$, and  $c_h^i$ are the effective corrections made by the platform. 
In order to quantify the performance of our reconstruction methods we use the corrections and interventions visible in the 5-minute dataset as the best proxy for the real corrections $c_h^i$. In general, we would like the estimated values $\hat c_h^i$ to reduce the following errors:

\begin{itemize}
    \item the fraction of \textit{lost corrections}, consisting of the fraction of real corrections that are lost in the reconstructed series;
    \[
    \frac{\sum_{h,i: c_h^i > \hat c_h^i} (c_h^i - \hat c_h^i) }{\sum_{h,i} (c_h^i)}
    \]
    \item the fraction the \textit{added corrections}, consisting of the corrections mis-added by the reconstruction methods, divided by the total real corrections;
    \[
    \frac{\sum_{h,i: \hat c_h^i >  c_h^i} (\hat c_h^i - c_h^i) }{\sum_{h,i} (c_h^i)}
    \]
    \item the fraction of \textit{lost interventions}, consisting of the fraction of interventions no longer visible in reconstructed series;
    \[
    \frac{\sum_{h,i: c_h^i = 0 } \mymathbb{1}_{ \hat c_h^i>0 }}{\sum_{h,i} \mymathbb{1}_{ c_h^i> 0}}
    \]
    \item the fraction of \textit{added interventions}, consisting of the number of mis-added interventions by reconstruction methods, divided by the total number of real interventions.
    \[
    \frac{\sum_{h,i: \hat  c_h^i = 0 } \mymathbb{1}_{ c_h^i>0 }}{\sum_{h,i} \mymathbb{1}_{ c_h^i> 0}}
    \]
\end{itemize}
When approximating the corrections with the negative views visible in the hourly collection, i.e. by taking $c_h^i = -\tilde v_h^i \mymathbb{1}_{\{v_h^i>0\}}$, we obtain the errors shown in Table~\ref{tab:tab1}. As the table shows, the loss of information is far from being negligible. We loose  66.31\% of the corrections and 60\% of the interventions. 

In the following we present two methods to reduce this information loss. The first, which we will refer to as the \textit{benchmark method}, is simpler and uses a heuristic. The second method, which we will refer to as the \textit{reconstruction method}, uses an XGBoost classifier \cite{xgboost} to improve the benchmark method and reduce information loss in the reconstructed series.

\begin{table}[!ht]
\begin{adjustwidth}{-1cm}{-1cm}
\begin{center} 

 \begin{tabular}[h!]{||c |c  c c||}
 \hline 
 &Hourly Aggregation  &Benchmark Method    &  Reconstruction Method  \\ [0.5ex]
 \hline\hline
 Lost Corrections & 66.31\% &   50.27 \%&  35.64\% \\ 
 \hline
 Added Corrections  & 0\% &   1.63\%&  4.93\% \\ 
 \hline
 Lost Interventions  & 60.00\% &   60.00 \%&  45.27\% \\ 
 \hline
 Added Interventions  & 0\% &   0 \%&  1.31\% \\ 
 \hline
\end{tabular}
\vspace{5pt}
\caption{\textbf{Validation of Reconstruction Method on the 5-minute-frequency dataset.} The table shows the loss of information with hourly aggregation and with the estimates done with the proposed Reconstruction Method and the Benchmark method.}
\label{tab:tab1}
\end{center}
\end{adjustwidth}

\end{table}

\subsection*{Benchmark Method}
\begin{figure}[h!]
    \centering
    \includegraphics[width=0.8\textwidth]{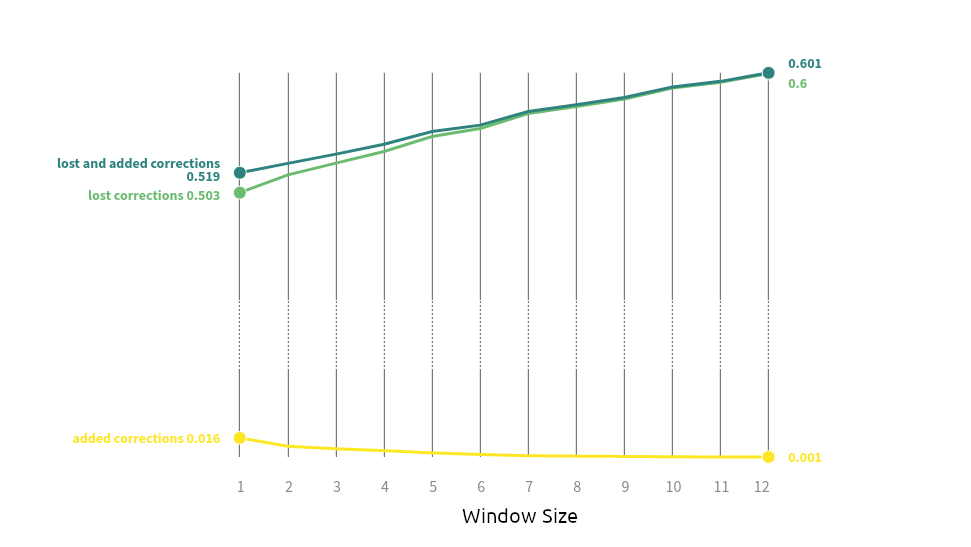}
    \caption{Error introduced by the benchmark method, varying the time window.}
    \label{fig:benchmark}
\end{figure}

As a benchmark method we propose the following heuristic: in hours with negative views, corrections are approximated by the number of negative views increased by the expected views at that hour. To estimate the expected views in a certain hour, we considered various quantities, such as the average and minimum number of views over a time window around the given hour. The time window considered are shown in Figure~\ref{fig:benchmark} and they should be intended symmetrical, e.g. a window size of two hour means that both the two hours preceding and following the hour of interest have been considered.
As shown in Figure~\ref{fig:benchmark} the errors considered in Table~\ref{tab:tab1} are minimal when the minimum over a one-hour time window is chosen as the approximation. Hence we approximate:
\begin{equation}
\label{eq:eq1}
   \hat c_h^i = (-v_{h}^i + \min (v_{h+1}^i, v_{h_1}^i )) \mymathbb{1}_{\{v_h^i>0\}}
\end{equation}
This benchmark method reduces the lost corrections from 66.31\% to 50.27\% by introducing only the 1.63\% of added corrections. \\

\subsection*{Reconstruction Method}

The benchmark method adjusts negative views to estimate the platform corrections, but it fails to detect correction events that have occurred when the observed views $\hat v_h^i$ are nonnegative. Hence we developed a method meant to detect anomalies in the views evolution and attribute them to concealed corrections. 
The method consists of an XGBoost classifier that can detect hours with unobserved corrections. 
Below we present how this classifier was trained and we analyze its performance.

We constructed the train dataset as follows:
\begin{itemize}
    \item each row represents one hour of our time-series;
    \item for each hour we extracted the evolution of views in the 12 hours before and after that hour and added them as features;
    \item we added as features the time of day (since corrections are roughly periodic and occur mostly between 4 and 6 p.m.) and the number of hours elapsed since publication.
\end{itemize}

\begin{figure}[h!]
    \centering
    \includegraphics[width=\textwidth]{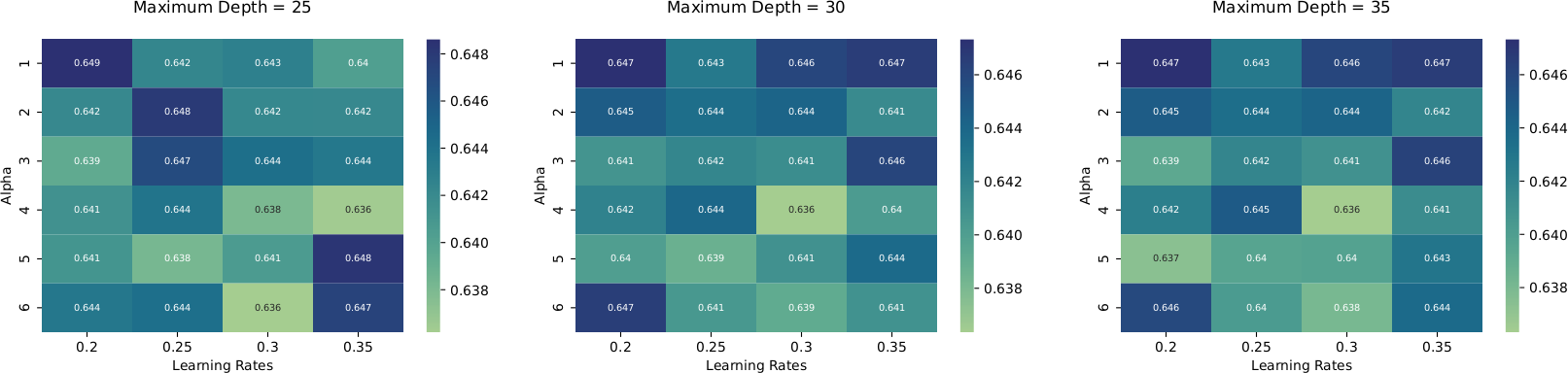}
    \caption{\textbf{XGBoost parameter tuning.} Performances in terms of F1 score associated with different combinations of parameters' values.}
    \label{fig:xgBoost}
\end{figure}

In order to do parameter tuning, we chose to optimize the F1 score, a metric defined as follows:
\[
\text{F1} = \frac{2}{\frac{1}{\text{precision}}\times\frac{1}{\text{recall}}}
\]
where \textit{precision} stands for the rate of true positives over all the samples classified as positive, while \textit{recall} stands for the rate of true positives over all the really positive samples in the data.
In this way we can limit the number of false positives and false negatives introduced by our classifier. We divided our dataset into a train set of 90075 observations and a test set of 24674 observations. 
We used the train set to perform a 5-fold cross validation to choose the optimal values of \textit{maximum depth} of the decision trees, the \textit{learning rate} and \textit{alpha} parameters of the XGBoost classifier. The performances in terms of F1 score associated with different combinations of values are shown in Figure~\ref{fig:xgBoost}. The best parameters, able to grant an F1 score equal to 0.649, are a maximum depth equal to 25, a learning rate equal to 0.2 and alpha equal to 1.

The classifier allows us to reconstruct which hours have had YouTube interventions without these being visible in the hourly aggregation. Once we identify the hours with interventions, we estimate the magnitude of the corrections by formula~\eqref{eq:eq1}. In this way, as shown in Table~\ref{tab:tab1}, we are able to reduce the lost corrections from 66.31\% to only 35.64\%.

\section{Conclusion}
The analysis of our data reveals that fake views are widespread and that most fake views are corrected relatively late in the life of YouTube videos. This observation rises concerns because of the key role of early popularity in determining overall visibility. Indeed, rich-get-richer dynamics have been repeatedly observed in the evolution of views counts on the platform \cite{clones}, and the total number of previous views has been credited as the most important predictor of future popularity \cite{clones} \cite{huberman} \cite{pinto}.
Furthermore, rewarding trendy contents with more visibility is a central feature of the recommendation system, which --according to YouTube developers-- aims at ``bootstrapping and propagating viral content'' \cite{covington_deep_2016}. While encouraging diffusion of trending videos, the recommendation system also constitutes the main source of views for most YouTube videos \cite{zhou_impact_2010} \cite{zhou2} and hence regulates the attention economy of the platform. Its suggestions are updated relatively quickly: according to Roth et al., for instance, two thirds of the suggestions are associated with a given video for less than two days \cite{roth:hal-02445546}. Such a fast refreshing of suggested videos, along with the massive impact of recommendation system, warrants concerns about delays in fake views correction. With a delay in correction, videos has handsomely enough time to be recognized as viral and thus be pushed to a wider audience than they would have achieved without illegitimate boosting.

Our work has brought to light the vast amount of channels and videos concerned by this phenomenon and its features in term of rhythms and frequency. We have also highlighted the existence of a positive correlation between illegitimate and real views. In the absence of first-hand data on fake views corrections, we cannot rule out the possibility that it is fake views that boost the popularity of some videos. Given the importance of the subject and the potential harm from the malfunctioning of the correction policy, our findings should --at the very least-- encourage YouTube to include in its API the number of corrected views for each videos, as well as their history. As we have shown, this information is crucial to investigate the alarming possibility that techniques of views inflation could be used to manipulate video visibility by triggering viral dynamics sustained by manual and algorithmic recommendation.

\section*{Acknowledgments}
A major acknowledgment goes to Yoan Dinkov and Preslav Nakov of the Qatar Computing Research Institute (QCRI) for their help in collecting YouTube data. A special thanks goes to Bilel Benbouzid for the precious discussions on the French YouTube landscape.

\section*{Declarations}

\begin{itemize}
\item  \textbf{Funding} \\
This research has been supported by CNRS through the 80 PRIME MITI project ''Disorders of Online Media'' (DOOM), by ANR through grant 19-P3IA-0003 and by the European Union’s Horizon 2020 project SoBigData++ (grant agreements No. 871042). 

\item \textbf{Availability of data and materials.}\\ 
Data is available on figshare at \cite{data1}, \cite{data2}, \cite{data3}.

\item \textbf{Code availability.}\\ Code is available at github \url{ https://github.com/mariacastaldo/Fake_views}.
\item \textbf{Authors' contributions.} \\
M.C.\ conceived and performed the research, collected and analyzed the data, discussed the results, wrote the manuscript. 
P.F., F.G., T.V.\ conceived the research, discussed the results, wrote the manuscript. 
All authors have read and approved the manuscript.
\end{itemize}









\bibliography{references}
\bibliographystyle{plainurl}


\end{document}